\documentclass[11pt,twoside]{article}

\usepackage{graphicx}

\usepackage{amssymb}      
\usepackage{savesym}      
\usepackage{amsmath}      
\savesymbol{iint}         
\usepackage{asp2010}
\restoresymbol{TXF}{iint} 
\usepackage{textcomp}

\resetcounters

\begin{document}

\resetcounters

\title{Power spectrum of gravitational waves from unbound compact binaries}
\author{Lorenzo De Vittori$^1$, Philippe Jetzer$^1$ and Antoine Klein$^2$ \affil{$^1$ University of Zurich, Institute for Theoretical Physics, Switzerland\\
$^2$ Montana State University, Departement of Physics, Bozeman, USA}}

\begin{abstract}
Unbound interacting compact binaries emit gravitational radiation in a wide frequency range.
Since short burst-like signals are expected in future detectors, such as LISA or advanced LIGO, it is interesting to study their energy spectrum and the position of the frequency peak.
Here we derive them for a system of massive objects interacting on hyperbolic orbits within the quadrupole approximation, following the work of Capozziello et al.
In particular, we focus on the derivation of an analytic formula for the energy spectrum of the emitted waves.
Within numerical approximation our formula is in agreement with the two known limiting cases: for the eccentricity $\varepsilon=1$,
the parabolic case, whose spectrum was computed by Berry and Gair, and the large $\varepsilon$ limit with the formula given by Turner.
\end{abstract}

\section{Theoretical framework}\label{sec:theory}

Since in the last years the Gravitational Waves (GWs) detection technology has improved very rapidly,
and it is believed that the precision we reached should enable their detection,
it is interesting to study the dynamics of typical systems and their emission of GWs and in particular
their frequency spectrum, in order to know at which wave-length range we should expect gravitational radiation.

For the cases of binary systems or spinning black holes on circular and elliptical orbits 
the resulting energy spectra have already been well studied, e.g. \cite{peters1,peters2}.
The energy spectrum for parabolic encounters has been computed either by direct integration along
unbound orbits by \cite{Turner} or more recently by taking the limit of the Peters and Mathews energy
spectrum for eccentric Keplerian binaries, see \cite{BerryGair}.

The emission of GWs from a system of massive objects interacting on hyperbolic
trajectories using the quadrupole approximation has been studied by \cite{Capozziello}
and analytic expressions for the total energy output derived.
However, the energy spectrum has been computed only for the large eccentricity ($\varepsilon \gg 1$) limit, see in \cite{Turner}.
Here we present the work done in the last year, \cite{Devittori}. We derive the energy spectrum for hyperbolic encounters for all values $\varepsilon \geq 1$
and we give an analytic expression for it in terms of Hankel functions.\\

GWs are solutions of the linearized field equations of General Relativity and the radiated 
power to leading order is given by Einstein's quadrupole formula, as follows
\begin{equation}\label{eq:poweremission}
 P = \frac{G}{45 \, c^5} \langle \dddot{D}_{ij} \, \dddot{D}_{ij} \rangle~,
\end{equation}
where we used as definition for the second moment tensors $M_{ij} := \frac{1}{c^2} \int T^{00} x_i x_j \, \mathrm{d}^3x$, and
for the quadrupole moment tensor $D_{ij} := 3 M_{ij} - \delta_{ij} M_{kk}$.
The quantity $M_{ij}$ depends on the trajectories of the involved masses, and can easily be computed for all type of Keplerian trajectories.
To compute the power spectrum, i.e. the amplitude of radiated power per unit frequency, requires a Fourier transform of equation (\ref{eq:poweremission}),
which is rather involved (for the elliptical case see e.g. \cite{Maggiore}), and we will derive it below for hyperbolic encounters.
The eccentricity $\varepsilon$ of the hyperbola is
\begin{equation} \label{eq:eccentricity}
 \varepsilon := \sqrt{1+2\,E\,L^2/\mu\,\alpha^2}~,
\end{equation}
where $E = \frac{1}{2}\mu\,v_0^2$ ($E$ is a conserved quantity for which we can take the energy at $t=-\infty$),
$v_0$ being the velocity of the incoming mass $m_1$ at infinity, the angular momentum $L=\mu\,b\,v_0$, the impact parameter $b$,
the reduced mass $\mu := \frac{m_1\,m_2}{m_1+m_2}$, the total mass $m:=m_1+m_2$, and the parameter $\alpha := G\,m\,\mu$.

Setting the angle of the incident body to $\varphi=0$ at initial time $t=-\infty$,
the radius of the trajectory as a function of the angle and as a function of time is given by
\begin{equation*}
 r(\varphi)=\frac{a\,(\varepsilon^2-1)}{1+\varepsilon \, \cos(\varphi-\varphi_0)} ~,\quad r(\xi)=a\,(\varepsilon\,\cosh\xi-1) 
\end{equation*}
with the time parametrized by $\xi$ through the relation $t(\xi)=\sqrt{\frac{\mu\,a^3}{\alpha}}\,(\varepsilon\,\sinh\xi-\xi)$,
where $\xi$ goes from $-\infty$ to $+\infty$.
Expressing this in Cartesian coordinates in the orbital plane, we finally get the equations for hyperbolic trajectories
\begin{equation}
 x(\xi)=a\,(\varepsilon-\cosh\xi)~,\qquad y(\xi)=a\,\sqrt{\varepsilon^2-1}\,\sinh\xi~.
\end{equation}

\section{Power spectrum of Gravitational waves from hyperbolic paths}\label{sec:powerspectrum}

\subsection{Power emitted per unit angle}\label{subsec:powerunitangle}

In \cite{Capozziello} the computation of the power emitted as a function of the angle, as well as the total energy emitted by the system has been already carried out.
They turn out to be:
\begin{equation}\label{eq:Delta_E}
 P(\varphi) = - \frac{32 \,G\,L^6 \mu^2}{45\,c^5\,b^8} f(\varphi,\varphi_0)~, \quad
 \Delta E = \frac{32\,G\,\mu^2\,v_0^5}{b\,c^5}\,F(\varphi_0)~,
\end{equation}
where for the factors $f(\varphi,\varphi_0)$ and $F(\varphi_0)$ one finds:
\begin{equation*}
 \begin{split}
  f(\varphi,\varphi_0) = \frac{\sin(\varphi_0 - \frac{\varphi}{2})^4 \; \sin(\frac{\varphi}{2})^4 }{\tan(\varphi_0)^2 \;
  \sin(\varphi_0)^6} \cdot \bigg(150 + 72 \cos(2 \varphi_0) + \\ + \; 66 \cos(2 (\varphi_0 - \varphi)) - 144 \; (\cos(2 \varphi_0 - \varphi) - \cos(\varphi))\bigg)~,
 \end{split}
\end{equation*}
\begin{equation*}
 \begin{split}
  F(\varphi_0) =& \frac{1}{720\,\tan^2\varphi_0 \, \sin^4 \varphi_0}\times[2628 \varphi_0 + 2328 \varphi_0 \cos 2\varphi_0\\
  +&144 \varphi_0 \cos4\varphi_0 - 1948\sin 2\varphi_0 - 301 \sin 4 \varphi_0]~.
 \end{split}
\end{equation*}
This means that the total radiated energy of the system can be determined knowing 
the parameters $b$ and $v_0$, and of course the reduced mass $\mu$.

\subsection{Power spectrum}\label{subsec:powerspec}

We compute now $P(\omega)$, the Fourier transform of $P(t)$, which describes the distribution of the amplitude
of the power emitted in form of GWs depending on the frequency.
In \cite{LL} and \cite{Longair} some hints are given when solving the analogous problem in electrodynamics.
The crucial idea is to use Parseval's theorem on the integration of Fourier transforms,
and then to express some quantities in terms of Hankel functions.
This allows to get in an easier way the function $P(\omega)$, for which we use the expression given in 
eq. (\ref{eq:poweremission})
\begin{equation}
 \begin{split}
 &\Delta E = \!\int \! P(t)\mathrm{d}t = \!\int \! P(\omega)\mathrm{d}\omega =
 -\frac{G}{45c^5} \!\int \! \textlangle \, \dddot{D}_{ij}(t) \, \dddot{D}_{ij}(t) \, \textrangle\,\mathrm{d}t=\\
 &-\frac{G}{45c^5} \int \! (|\widehat{\dddot{D}_{11}}(\omega)|^2 + |\widehat{\dddot{D}_{22}}(\omega)|^2
 + 2 |\widehat{\dddot{D}_{12}}(\omega)|^2 + |\widehat{\dddot{D}_{33}}(\omega)|^2 )\,\mathrm{d}\omega~,
 \end{split}
\end{equation}
where $\widehat{\dddot{D}_{ij}}(\omega)$ is the Fourier transform of $\dddot{D}_{ij}(t)$.
It is easy to see that the last equation represents the total amount of energy dissipated in the encounter.
Therefore, the integrand in the last line has to be equal to the power dissipated per unit frequency $P(\omega)~$:
\begin{equation}\label{eq:P_omega}
 P(\omega)= -\frac{G}{45c^5} \, \big( |\widehat{\dddot{D}_{11}}(\omega)|^2 + |\widehat{\dddot{D}_{22}}(\omega)|^2 + 2 |\widehat{\dddot{D}_{12}}(\omega)|^2 + |\widehat{\dddot{D}_{33}}(\omega)|^2 \big)~.
\end{equation}

As next, we need to compute the $\widehat{\dddot{D}_{ij}}(\omega)$, take the square their norm and add them together, which yields the power spectrum.
Computing the $D_{ij}$ explicitly - keeping in mind that we use the time parametrization $t(\xi) = \sqrt{\mu\,a^3/\alpha}\,(\varepsilon\,\sinh\xi-\xi)$ - we get:
\begin{equation*}
 \begin{split}
 D_{11}(t) \sim ((3-\varepsilon^2) \cosh 2\xi - 8\,\varepsilon \cosh \xi)~,\quad &D_{22}(t) \sim (4\,\varepsilon \cosh \xi + (2\,\varepsilon^2-3) \cosh 2\xi)~,\\
 D_{33}(t) \sim (4\,\varepsilon \cosh \xi + \varepsilon^2 \cosh 2\xi)~,\quad \,\qquad &D_{12}(t) \sim (2 \, \varepsilon \sinh \xi - \sinh 2\xi )~.
 \end{split}
\end{equation*}
The Fourier transform of the third derivatives of $D_{ij}(t)$ is given by $\widehat{\dddot{D}_{ij}}(\omega) = i \omega^3 \, \widehat{D_{ij}}(\omega)$,
thus we have just to compute $\widehat{D_{ij}}(\omega)$. We can closely follow the calculations performed in \cite{LL},
where the similar problem in electrodynamics of the emitted power spectrum for scattering charged particles on hyperbolic orbits is treated.
In particular the following Fourier transforms are used (for their derivation see Appendix A in \cite{Devittori}).
\begin{equation} \label{eq:HankelRelations}
  \widehat{\sinh \xi} = - \frac{\pi}{\omega \varepsilon} H_{i \nu}^{(1)}(i \nu \varepsilon) \; , \quad \widehat{\cosh \xi} = - \frac{\pi}{\omega} H_{i \nu}^{(1)}\textquotesingle(i \nu \varepsilon)~,
\end{equation}
\begin{equation} \label{eq:HankelDerivative}
 H_{\tilde{\alpha}}^{(1)}\textquotesingle(x) = \frac{1}{2} (H_{\tilde{\alpha}-1}^{(1)}(x)-H_{\tilde{\alpha}+1}^{(1)}(x))~,
\end{equation}
where $H_{\tilde{\alpha}}^{(1)}(x)$ is the Hankel function of the first kind of order $\tilde{\alpha}$, and where $\nu$ is defined as $\nu := \omega \sqrt{\mu a^3/\alpha}$.\\
Taking the $D_{ij}(t)$ from above we get
\begin{equation*}
\begin{split}
 \widehat{D_{11}}(\omega) &= \frac{a^2\,m\,\pi}{4\,\omega}\,[16\,\varepsilon\,H_{i \nu}^{(1)}\textquotesingle(i\nu\varepsilon)+(\varepsilon^2-3)\,H_{i\nu}^{(1)}\textquotesingle(i\nu\varepsilon/2)]~,\\
 \widehat{D_{22}}(\omega) &= \frac{a^2\,m\,\pi}{4\,\omega}\,[(3-2\,\varepsilon^2)\,H_{i \nu}^{(1)}\textquotesingle(i\nu\varepsilon/2)-8\,\varepsilon\,H_{i\nu}^{(1)}\textquotesingle(i\nu\varepsilon)]~,\\
 \widehat{D_{33}}(\omega) &= \frac{a^2\,m\,\pi}{4\,\omega}\,[8\,\varepsilon\,H_{i \nu}^{(1)}\textquotesingle(i\nu\varepsilon)+\varepsilon^2\,H_{i\nu}^{(1)}\textquotesingle(i\nu\varepsilon/2)]~,\\
 \widehat{D_{12}}(\omega) &= \frac{3\,a^2\,m\,\pi}{4\,\omega\,\varepsilon}\,\sqrt{\varepsilon^2-1}\,[H_{i \nu}^{(1)}(i\nu\varepsilon/2)-4\,\varepsilon\,H_{i\nu}^{(1)}(i\nu\varepsilon)]~.
\end{split}
\end{equation*}
Inserting this result into eq. (\ref{eq:P_omega}), and using the formula for the Fourier transform of the third derivative, we get the power spectrum of the gravitational wave emission for hyperbolic encounters
\vspace{0.1in}
\begin{equation} \label{eq:P_omega_mio}
 P(\omega)=- \frac{G\,a^4\,m^2\,\pi^2}{720\,c^5} \; \omega^4 \; F_{\varepsilon}(\omega)~,\vspace{0.1in}
\end{equation}
where the function $F_{\varepsilon}(\omega)$ turns out to be
\begin{align*}
 \begin{split}
 |[16\,\varepsilon\,H_{i \nu}^{(1)}\textquotesingle(i\nu\varepsilon)+(\varepsilon^2-3)\,H_{i\nu}^{(1)}\textquotesingle(i\nu\varepsilon/2)]|\,^2 +\,&
 |[(3-2\,\varepsilon^2)\,H_{i \nu}^{(1)}\textquotesingle(i\nu\varepsilon/2)-8\,\varepsilon\,H_{i\nu}^{(1)}\textquotesingle(i\nu\varepsilon)]|\,^2\\
 +\;|[8\,\varepsilon\,H_{i \nu}^{(1)}\textquotesingle(i\nu\varepsilon)+\varepsilon^2\,H_{i\nu}^{(1)}\textquotesingle(i\nu\varepsilon/2)]|\,^2+\,&
 \frac{9\,(\varepsilon^2-1)}{\varepsilon^2}\,|[H_{i \nu}^{(1)}(i\nu\varepsilon/2)-4\,\varepsilon\,H_{i\nu}^{(1)}(i\nu\varepsilon)]|\,^2~.\\
 \end{split}
\end{align*}

In Fig. 1 the function $\omega^4 \, F_{\varepsilon}(\omega)$ is plotted for some some values of $\varepsilon$:
this is the frequency power spectrum of gravitational radiation emitted by an hyperbolic encounter.
Unfortunately the expression for $F_{\varepsilon}(\omega)$ is rather complicated and we could not find an analytical way to simplify it.
We thus made some numerical tests to check its validity and clearly the integral of (\ref{eq:P_omega_mio}) has to be equal to $\Delta E$ in (\ref{eq:Delta_E}),
which was obtained by integrating over the power emitted per unit frequency, i.e. $\int_0^{\infty} \!P(\omega) \,\mathrm{d}\omega = \Delta E$.
\begin{figure}[!htb]\vspace{0.08in}
 \begin{center}
  \includegraphics[width=0.8\textwidth]{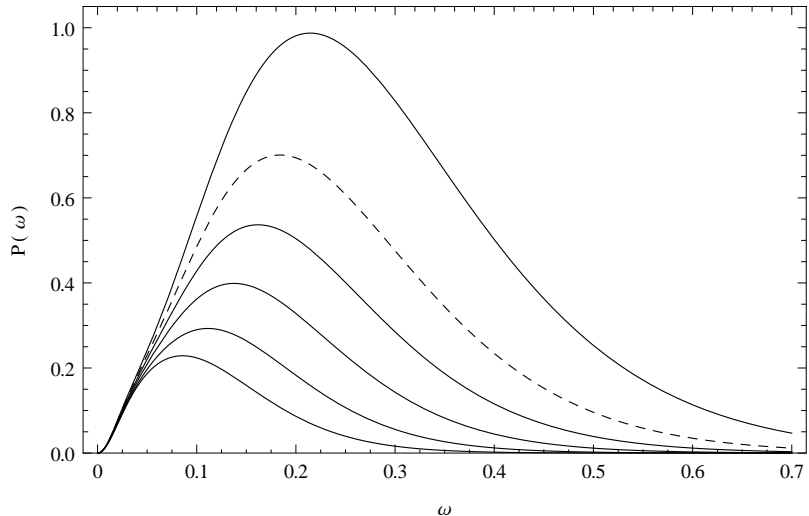}
  \label{fig:Powerspectrum}
  \caption{The frequency power spectrum of gravitational radiation emitted by an hyperbolic encounter. On the $x$-axis we have the angular frequency $\omega$ expressed in mHz units,
  whereas on the $y$-axis the amplitude of $P(\omega)$ is normalized to the maximum value of the $\varepsilon\sim2.5$ case.
  These are the expected emissions generated by a system of two supermassive black holes with $m = 10^7 M_{\odot}$, impact parameter $b=10$ AU, and different relative velocities.
  With lower velocities the interactions are stronger and the eccentricity decreases.
  These spectra, in order from the highest to the lowest, represent systems with $v_0 = 3.4 \times 10^7$ m/s ($\varepsilon \sim 2.5$), $v_0 = 3.5 \times 10^7$ m/s ($\varepsilon\sim3$),
  $v_0 = 3.6 \times 10^7$ m/s ($\varepsilon\sim3.1$), $v_0 = 3.75 \times 10^7$ m/s ($\varepsilon\sim3.4$), $v_0 = 4 \times 10^7$ m/s ($\varepsilon\sim3.8$),
  $v_0 = 4.5 \times 10^7$ m/s ($\varepsilon\sim4.7$), respectively.
  In particular the case with $\varepsilon\sim3$ (plotted with the dashed line) is discussed in the conclusions.
  As one can see, for higher eccentricities the peak frequency slowly decreases. This is only true for values of $v_0$ up to $\sim 6 \times 10^7$ m/s,
  whereas above it increases again. Moreover, decreasing the mass or increasing the impact parameter changes the eccentricity as well.
  We should be able to detect incoming waves in that range e.g. with eLISA, since the peak at $\sim 0.2$ mHz fits in its observable band.
  For a more detailed discussion see Sec. 3 and e.g. \cite{LISA}.}
 \end{center}
\end{figure}

We have checked the validity of this equality for different sets of values, comparable to those used in \cite{Capozziello},
e.g. $b=1\textmd{AU}$, $v_0 = 200$ km/s, and $m_{1,2}=1.4 \;\textmd{M}_{\odot}$, or similar.
For all of these sets we got agreement within numerical accuracy.

More interesting is the case where the eccentricity approaches $\varepsilon=1$.
According to eq. (\ref{eq:eccentricity}) this is the case e.g. with the set of initial conditions $b=2$ AU, $v_0 = 6.4$ km/s and $m_{1,2}=1.4\,M_{\odot}$.
Since this is a limit case for a parabolic trajectory, we can directly compare our result with the one studied by \cite{BerryGair},
and indeed they coincide, within numerical accuracy.
For a discussion about the feasibility of an analytical comparison see Appendix B in \cite{Devittori}.

Finally, we turn to the large  $\varepsilon$ limit and compare our result with the one given in \cite{Turner}
and \cite{WW}.
The expression for the total energy emitted during an hyperbolic interaction is written in \cite{Turner} as:
\begin{equation}\label{eq:E_Turner}
 \Delta E_T = \frac{8}{15} \frac{G^{7/2}}{c^5} \frac{m^{1/2}\,m_1^2\,m_2^2}{r_{min}^{7/2}} ~g(\varepsilon)~, \quad\textmd{where for }\varepsilon\rightarrow\infty:\quad g(\varepsilon) \sim \frac{37 \pi}{8}\sqrt{\varepsilon} \;+ \,\mathcal{O}(\,\varepsilon^{-1/2})~,
\end{equation}
which also agrees with the result of \cite{WW}.

Comparing our total energy from the quadrupole approximation, eq. (\ref{eq:Delta_E}), with the expression for the energy $\Delta E_T$ (\ref{eq:E_Turner}) by \cite{Turner}
valid in the large $\varepsilon$ limit, we see that they coincide for large eccentricities,
having e.g. a $1\%$ difference after $\varepsilon=100$, and a $5\%$ difference after $\varepsilon=20$.
For a more detailed discussion about these comparisons with previous results, see our full work \cite{Devittori}.

\section{Conclusions}\label{sec:conclusions}

Short gravitational wave burst-like signals are expected in the data stream of detectors.
Although these signals will likely be too short to allow us to measure the parameters of the emitting system accurately,
the results presented in this paper could be used to get a rough estimate of these parameters,
by observing the position of the peak, the amount of energy released and the timescale of the interaction.

Given the knowledge of the power spectrum we can easily see which kind of hyperbolic encounters could generate gravitational waves detectable e.g. with eLISA, advanced LIGO or advanced VIRGO.
Measurements from unbound interactions with ground-based detectors could in principle be possible, though the energy emitted at e.g. $\pm 200$ Hz
is below the minimum threshold for advanced LIGO or advanced VIRGO, making detections unlikely but not impossible.
The space-based interferometer instead is expected to cover frequencies ranging from $0.03$ mHz up to $1$ Hz, see e.g. \cite{LISA}, where the interactions could release more energy.

An unbounded collision between two intermediate-mass black holes, let's say of $10^3 M_{\odot}$ each, with an encounter velocity of $2000$ km/s at a distance of $1$ AU, would generate,
according to our eq. (\ref{eq:P_omega_mio}), a frequency spectrum with peak around $0.04$ mHz, with $80\%$ of the emission in the range between $0.01$ and $0.07$ mHz,
i.e. in the lower range limit of eLISA.
Another possible example of measurable impact would be an encounter between two supermassive black holes with mass, e.g., comparable to the expected mass of Sagittarius A*,
the black hole believed to be at the center of our galaxy, i.e. $\sim 10^7 M_{\odot}$.
With a distance of some AU, and a high velocity (we want to exclude the bounded case) of tens of thousands km/s, such a collision would generate
an energy spectrum with peak at $\sim 0.2$ mHz with $80\%$ between $0.03$ and $0.37$ mHz, thus in the observable range of eLISA.
(Its energy spectrum is plotted with a dashed line in Fig. 1.)
Estimates for the rate of such events have been considered e.g. in \cite{cl}. They consider e.g. typical compact stellar cluster around the Galactic Center,
and expect an event rate of $10^{-3}$ up to unity per year, depending on the radius of the object and the amount of such clusters in the near region.

We believe that with the energy spectrum found here one should be able to classify the different encounters depending on t different encounters depending on the detected shape,
and therefore get a better insight into the map of our galaxy or the near universe.\\

\begin{acknowledgments}
We thank N. Straumann for useful discussions and for bringing to our attention the relevant treatment of the hyperbolic problem in electrodynamics in Landau \& Lifschitz.
We also thank L. Blanchet for his encouragement and for pointing out the possibility of treating the same problem in another way.
Finally, we would also like to thank C. Berry for helping clarifying some details.
\end{acknowledgments}

\bibliography{MyBibli}
\bibliographystyle{asp2010}

\end{document}